**Impact of Criterion-Based Reflection on Prospective Physics Teachers' Perceptions of ChatGPT-Generated Content**

Farahnaz Sadidi* and Thomas Prestel

Chair of Physics Education, TU Dresden, 01069 Dresden, Germany

**ABSTRACT**. ChatGPT has significantly shaped digital transformation discussions. Its widespread testing and ongoing optimization highlight the need to assess its capabilities and encourage critical reflection. This study explores how students' critical reflection on ChatGPT-generated content impacts their perceptions of its answer quality and further use. Involving 39 prospective physics teachers, the study assessed their evaluations of ChatGPT's answers to didactical tasks using predefined criteria, in this paper referred as the criterion-based evaluation approach. Pre- and post-questionnaires, with a 5-point ranking scale and open-ended questions, evaluated students' perceptions of ChatGPT's helpfulness and quality. Results showed that critical reflection shifted students' perception of answer quality and increased their awareness of ChatGPT's limitations. Three perspectives on ChatGPT's further use emerged: extreme positive, extreme negative, and balanced. The extreme positive perspective indicated confirmation bias, where the positive prior experiences of students with ChatGPT influenced their evaluations. The findings highlight the need to foster critical thinking and media literacy in teacher training programs to help educators effectively integrate such tools.

## I. INTRODUCTION

The release of ChatGPT in November 2022 generated significant media interest. The potential of a universal chatbot that responds to any posed question or task ("prompt") is very promising and has garnered considerable attention.

Researchers subsequently evaluated the bot's performance across various disciplines, including physics [1]. ChatGPT3.5 already delivers good results in advanced physics tests, with later versions performing almost like physics experts [2]. Improved prompting strategies have also been shown to enhance the chatbot's output [3,4].

The availability of such a bot raises concerns about exam formats in high schools and universities, particularly those focused on simple knowledge reproduction, especially in online settings [5]. However, ChatGPT struggles with "higher-order thinking problems", making it essential to develop tasks involving reflection and critical thinking to assess student competencies [6,7].

In physics education, while ChatGPT can support prospective teachers in developing tasks for high school students, critical reflection is crucial to ensure that the generated content is clear, specific, and contextually relevant [8].

Studies on student interactions with ChatGPT reveal a troubling overreliance on AI, as students often accepted (partly) incorrect solutions from ChatGPT as correct, even in their areas of expertise [9].

Considering the potential and limitations of ChatGPT on the one hand and the challenges students face when using this tool on the other hand shows how important it is to promote critical thinking when integrating ChatGPT into education [8].

This issue is particularly relevant in teacher training programs, where media literacy is a key competency [10]. Teachers are expected to critically evaluate media, considering its technological, social, and application-related aspects. As AI becomes increasingly important, a new set of skills, known as AI literacy, has emerged [11], with assessment tools already developed [12]. However, for initial interactions with AI chatbots and in line with the broader goals of teacher training programs, we continue to use the overarching term "media literacy". In media literacy, the technological perspective involves understanding the functionality and technical aspects of tools like ChatGPT. The social perspective focuses on critically assessing the impact of such tools on social interactions. The application-related perspective emphasizes effectively integrating these tools into teaching and learning processes. To develop media literacy and critically reflect on media use, students must have critical thinking skills to apply relevant criteria for evaluating content and explanations effectively.

Studies on students' reasoning skills in physics often highlight deficiencies in their use of precise terminologies and their ability to provide coherent explanations [13]. These shortcomings become even more evident when students engage with AI-generated content [9]. This study therefore introduces a "criterion-based evaluation approach" to help students effectively evaluate ChatGPT-generated content using clear evaluation criteria.

To explore ChatGPT's potential in fostering critical thinking and media literacy, it is integrated into a didactical seminar for prospective physics teachers. Since the tool was relatively new at the time of the study, in spring 2023, students' interactions with it

---

were uncertain. This investigation aims to shed light on its potential benefits in broader educational contexts.

The research focuses on the **overarching question**: "What is the effect of a criterion-based evaluation approach on students' perceptions of ChatGPT's quality and its further use?"

## II. METHODOLOGY AND RESEARCH QUESTIONS

The study employs a criterion-based evaluation approach, grounded in the Critical Thinking framework [14], to enhance students' critical reflection in evaluating ChatGPT3.5-generated content. The approach emphasizes verbal reasoning skills, highlighting the need to avoid misleading information by identifying vague or ambiguous terms and using precise language. In physics education, explanations must meet criteria such as clarity, correctness, appropriateness, and precision to deepen students' understanding [15]. These criteria are also crucial for developing media literacy, enabling teachers to critically assess the use of technology.

The study targets physics teacher students, engaging them in evaluating ChatGPT-generated content in didactical tasks. By using explicit evaluation criteria, the aim is to investigate how the students assess the quality of ChatGPT's output and how this impacts their perceptions of the tool's helpfulness and its answer quality.

The research is designed as an exploratory study to gain insights into how a structured evaluation process can shape students' reasoning and critical reflection when interacting with AI tools. To address this, the following research questions (RQs) are asked:

**RQ1:** How do prospective physics teachers assess the quality of ChatGPT3.5-generated content for tasks like "developing context-oriented teaching in acoustics" and "formulating learning goals"?
This question explores how physics teacher students evaluate ChatGPT's performance in developing teaching materials and learning objectives, which are essential for effective instructional planning. Reflecting on whether ChatGPT's content aligns with pedagogical goals helps develop students' media literacy, particularly in its practical application.

**RQ2:** How do prospective physics teachers evaluate the quality of ChatGPT3.5-generated content for specific didactical tasks using criteria such as clarity, correctness, appropriateness, and precision?
This question examines how effectively students apply the given criteria to AI-generated content and how these criteria help them engage in critical reflection.

**RQ3:** How does engaging prospective physics teachers in evaluating ChatGPT3.5-generated content impact their perception of ChatGPT's helpfulness and its answer quality?
This question investigates how participating in the criterion-based evaluation approach influences students' overall view of ChatGPT. It looks at whether the evaluation activity changes their perception of the tool's helpfulness and the quality of its answers over time.

**RQ4:** How do prospective physics teachers perceive and interpret the functionality of ChatGPT?
This question explores students' understanding of the technical aspects of ChatGPT, including its functionality. Observing their challenges and successes in the interpretation of this tool provides insight into the development of their media literacy, particularly from a technological perspective.

## III. STUDY DESIGN

The study was conducted as exploratory research within the study course of prospective physics teachers (see the flowchart of the study, Fig. 1). To investigate prospective teacher perceptions of ChatGPT generated content, the students were handed out pre- and post-questionnaires using a 5-point ranking scale, complemented with open-ended questions.

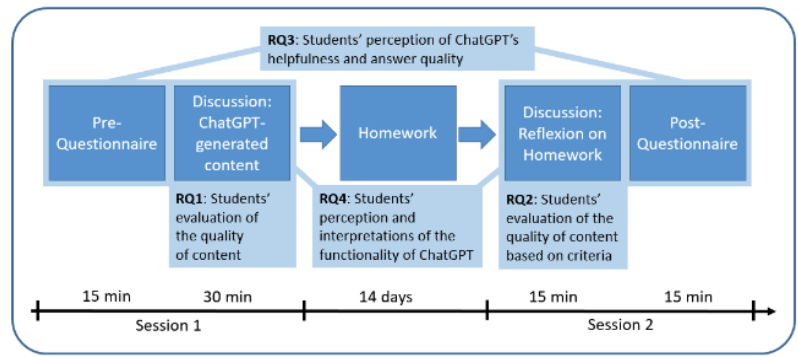

FIG 1. Flowchart of the study. The time estimates in minutes are approximate.

In the first session of the study, before encountering ChatGPT-generated content, students rated their estimation of the helpfulness and quality of ChatGPT3.5-generated content and provided explanations for their ratings. Subsequently, they were given ChatGPT's answer to a didactical task, for which they had to be prepared at home. The task focused on the development of a context-oriented teaching sequence for grade 10 students in acoustics and defining the learning goals.

They evaluated the ChatGPT content in a group discussion, which was audio recorded. For some of the students, this was their first contact with ChatGPT-generated content.

After this session, the students got homework to complete further didactical tasks using ChatGPT. The task was developing teaching-learning sequences on energy and power for 7th-grade students. Here, students were asked to evaluate the quality of the ChatGPT's answer specifically by using given criteria like clarity, correctness, appropriateness, and precision. For each criterion, a definition was given to clarify the aspects of evaluation to students (see Table 1). For example, by evaluating the clarity of explanation of ChatGPT, students should assess whether applied terminologies fit the context, whether they need further specification and whether the assumptions were mentioned, if applicable.

In the second session, students then re-ranked the helpfulness and quality of the ChatGPT answer. They also discussed their evaluation results in groups. Moreover, they discussed their attitude towards the impact of ChatGPT on teaching and learning in schools and reflected on the further use of the tool. These discussions were also audio recorded.

### III. PARTICIPANTS

Thirty-nine German physics teacher students participated in the study, including nine females. The students, with an average age of 22, were in their third year of the physics teacher training program. Of the participants, 24 had prior experience using ChatGPT3.5 – 21 males and 3 females – while 15 had no previous experience with the tool.

### IV. RESULTS

The qualitative analysis of students' responses to the pre- and post-questionnaires, along with their group discussions, addressed the study's research questions. Additionally, students' rankings of answer quality and the helpfulness of ChatGPT provided quantitative insights. However, recognizing the study's limitation of a small sample size (N=39), we integrated both quantitative and qualitative perspectives in our discussion of the results.

#### A. Students' evaluation of content quality from a didactical perspective

When students engaged with ChatGPT's answers in specific didactical tasks – developing context-oriented teaching in acoustics for grade 10 and formulating learning goals – data analysis showed consistent findings.

All students recognized ChatGPT's difficulty in using measurable operational verbs for learning goals, a crucial aspect of German university didactics courses. For instance, they criticized the learning goal formulated by ChatGPT: "The students understand that sound is a form of mechanical wave.", noting that the verb "understand" is not directly measurable. In contrast, operationalized verbs emphasize measurable outcomes, such as "explain" or "classify".

However, 38 out of 39 students did not recognize ChatGPT's lack of context consideration in task formulation. A traditional example of context-oriented teaching in acoustics is using the "Tacoma Bridge Collapse" [16] as context to teach the concept of "Resonance". In contrast, ChatGPT formulated the tasks such as "Explain the meaning of amplitude, frequency and wavelength for acoustic waves.", which is general and lacks context. This shows that students struggle with developing context-oriented teaching, despite this being covered in their introductory physics didactics course.

TABLE I. Criteria for evaluation of ChatGPT3.5- generated content and their definitions.

| Criterion | Definition |
|---|---|
| Clarity | The explanation includes the terminologies that fit the context and do not need further specification. Moreover, in the explanation, the assumptions are mentioned, when applicable. |
| Correctness | The explanation includes correct information based on content knowledge. |
| Appropriateness | The explanation fits students' grade and prior knowledge. |
| Precision | The explanation does not need further specification. |

## B. Students' evaluation of content quality based on criteria

During discussions about their experiences using ChatGPT to complete didactical tasks on developing teaching-learning sequences on energy and power for 7th-grade students, most students criticized the responses for lacking clarity, correctness, appropriateness, and precision. They applied the given criteria to highlight the deficiencies in ChatGPT's answer quality.

*Clarity*: Students criticized the clarity of ChatGPT's answer, noting it explained terminologies using other technical terms, such as defining energy as "Energy is the ability to perform work.", which is critical in didactical context.

*Correctness*: Students criticized the correctness of the ChatGPT's answer arguing its failure in providing an answer that is correct based on the content knowledge. Specifically, when addressing high school students' misconception in the concept of energy and power with ChatGPT, students realized that some mentioned misconceptions are nonsense. They also realized that the list of mentioned misconceptions was very incomplete and realized referring to the literature to have a logical and broader picture of students' misconceptions necessary.

S14: *"... about the student' misconceptions, I don't know, he [ChatGPT] calls them students' misconceptions, they can be there, but I find them so incredibly easy to refute. Because, for example, energy only exists in the form of electricity, he [ChatGPT] tells me. So, children can think that, but I think as soon as you start talking about thermal energy or kinetic energy, every pupil knows that it's complete nonsense ..."*

*Appropriateness of the answer to the students' level*: Students realized ChatGPT's weakness in connecting the presented answer to the students' grade since this aspect in education plays a very important role. They criticized the appropriateness of the answer to the grade 7th students, arguing its failure in considering students' prior knowledge and its ignorance of the use of appropriate terminologies for targeted school level.

S07: *"So in my case, it almost completely neglected the class level, from what I noticed. It gave me more general suggestions. And it didn't go into the explanations of terms for grade 7 either. I found that quite surprising ..."*

*Precision*: They criticized the precision of answers by arguing that the answer is superficial and also discussing explicitly the missing parts in the definition and explanations. Most of the students criticized ChatGPT's ability on sketching the teaching-learning sequences for teaching energy and power, arguing that although its answer included some suggestions, it did not build on each other to fulfill the unity and logical structure of the teaching-learning sequences.

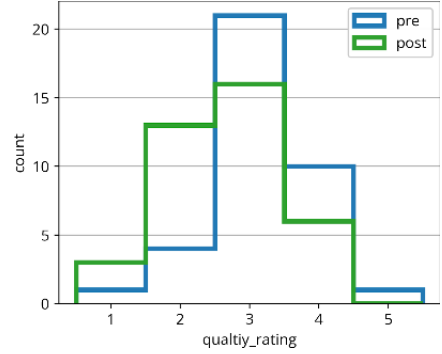

FIG 2. Distribution of quality ratings of ChatGPT's answers of the students in pre and post questionnaire.

## C. Students' perceptions of ChatGPT's answer quality and helpfulness

The histogram in Fig. 2 displays students' rankings of ChatGPT's answer quality on a 5-point ranking scale, comparing responses from the pre- to post-questionnaire. The results show a slight shift towards a more negative perception of its answer quality. This shift is also reflected in the mean value, which decreased from 3.16 to 2.66 (see Table II).

Similar to the answer quality, the mean value of students' perception of ChatGPT's helpfulness also decreased. Although this change is not statistically significant, a shift in students' perceptions of both ChatGPT's answer quality and its helpfulness was observed, reflecting initial higher expectations followed by practical disillusion.

Given the study's limitation of a small sample size, discussing the results from a qualitative perspective adds more plausibility. Visualizing students' ratings of ChatGPT's helpfulness in the pre- and post-questionnaires (Fig. 3) is particularly useful. The data visualization highlights notable cases of changes in students' rankings, both increases and decreases, in ChatGPT's helpfulness and answer quality.

TABLE II. Mean value and standard deviation of students' rating of the ChatGPT's answer helpfulness and its quality before (pre) and after (post) engaging with specific didactical tasks.

|  | **Pre** Mean (SD) | **Post** Mean (SD) |
|---|---|---|
| Quality | 3.16 (.76) | 2.66 (.85) |
| Helpfulness | 3.31 (.95) | 3.08 (1.05) |

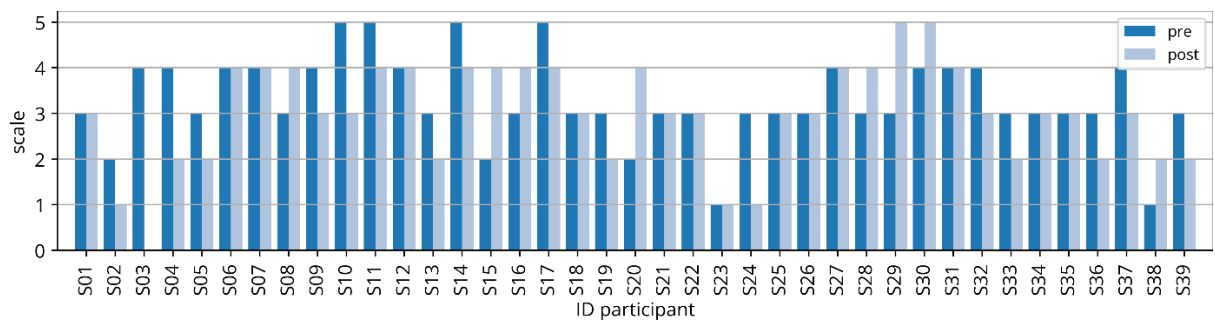

FIG 3. Visualization of rating of ChatGPT's helpfulness before (pre) and after (post) engaging with didactical tasks (Statement: ChatGPT is helpful: 1: completely disagree … 5 completely agree).

For instance, a 2-point decrease in helpfulness was observed for students S04, S10, and S24, while a 2-point increase was noted for students S15, S20, and S29 (see Fig. 3).
The following section presents the findings from the qualitative analysis of students' written answers and discussions on both aspects.

*1. Positive perceptions of helpfulness*

Qualitative analysis of students' answers in these cases showed that students who perceive ChatGPT helpful discuss its helpfulness in both general and physics contexts. In general context, its usefulness for brainstorming, inspiration and everyday things like writing thank-you cards is perceived positively. In the physics context (using ChatGPT in homework to answer the physics didactical tasks like developing teaching-learning sequences on energy and power for 7th-grade students), they perceived ChatGPT's ability in providing examples when explaining the physics terminologies positively. Moreover, the fact that ChatGPT provided the step-by-step instruction in teaching the concepts of energy and power is perceived positively.
Regarding the helpfulness of ChatGPT, students often highlight its ability to provide quick answers as a key indicator, as evident in the following students' statements:

S29 (in written form): "*ChatGPT quickly provides a clear answer; in literature sources, the essential answers to the questions are more hidden and embedded in a context.*" Or

S05-S06 (in discussion): "*That's actually quite good, because you don't have to struggle to find everything on the Internet in order to understand it and then perhaps have it explained to you in a shitty way, but you can actually ask the part and there's a relatively high probability that it will give you a proper answer that is probably also halfway understandable. So that's my experience so far.*"

*2. Critical evaluations of helpfulness*

On the other hand, students, who criticized the helpfulness of the ChatGPT's answer in the post-questionnaire, gave more weight to the criteria of clarity, correctness, appropriateness, and precision offered to them during the homework. They questioned the precision of ChatGPT's answers, since ChatGPT in answering their questions explained the terminologies by using other terminologies which themselves needed a definition. They also recognized errors in ChatGPT's answers and criticized its superficial answers. For example, S24 discussed:

"*I thought to myself afterwards that I wouldn't use ChatGPT. I would rate it a 5* [failing in German grade system], *if not a 6* [which is the worst grade in German grade system]. *It was so incredibly unhelpful. It was just phrases one after the other. And then sometimes things that were repetitive. For example, the student's idea that power is only associated with physical effort. And then later mentioned that power is linked to physical effort. Yes, that thing. My God.*"

*3. Critical evaluations of answer quality*

Using criteria, students criticized the quality of ChatGPT's answer in the didactical task and argued that its answers

- lack the basis content knowledge for example energy conservation laws
- include false information about the students' misconception in energy and power
- include inconsistency among the mentioned students' misconception and proposed teaching sequences
- are not appropriate for 7th-grade students

Following example shows the change of students' perspective before and after engaging with the task about the answer quality of ChatGPT. In the pre-questionnaire, the student ranked the quality of ChatGPT's answers as a 4. However, after completing the didactical task using specific criteria, the student downgraded the ranking to a 2.

S14 (in pre-questionnaire ranked the answer quality as a 4 [I agree]): *"I wrote a paper with ChatGPT, I had to correct all the texts a bit, but in the end it was enough for a 2.0 [good grade in German grade system]."*

S14 (in post-questionnaire ranked the answer quality as a 2 [I disagree]): *"The quality always depends on the question. The more complex the question, the worse the answer."*

### *4. Confirmation bias*

Analysis of students' responses who had a positive change (+1) in their perception of ChatGPT's answer quality from pre- to post-questionnaire revealed an interesting trend. While these students used criteria to critically evaluate ChatGPT's answers, often describing them as superficial, incomplete, and imprecise, they were surprised and impressed that ChatGPT provided examples to explain terminologies like work and power. This trend was also observed among students who rated ChatGPT's answer quality as a 4 in both the pre- and post-questionnaires, showing no change in their perception of its quality. While these students recognized the shortcomings of ChatGPT's answers based on given criteria, they were impressed by its ability to handle general tasks and suggest teaching-learning sequences. This admiration hindered their critical evaluation of the answer quality, a phenomenon explained by confirmation bias [17]. Confirmation bias is the tendency to favor evidence that supports existing beliefs while overlooking contradictory information. In this case, the students' positive impression of ChatGPT's general capabilities likely increased the probability that they would discount its shortcomings, leading them to overestimate its answer quality. This tendency was mostly observed among students who had prior experience with ChatGPT. The following student's statement serves as an example of confirmation bias:

S34: *"On the subject of clarity, I think the wording is clear, it's just **superficial, but you understand what the AI means**."*

### D. Students' interpretation of ChatGPT's functionality

Students had no specific task to comment on the functionality of ChatGPT, but while completing the given tasks, they occasionally expressed implicit ideas about its functionality through both discussion and writing. An understanding of ChatGPT as a database was the most common idea of its functionality, mentioned by many students. Examples are: "It's just a database" (S06) or "... major competence of the database" (S22). Others interpret ChatGPT as a search engine "like google" or an index search, when mentioning "He looks up what is relevant in class 7" (S15). In the last statement ChatGPT is additionally attributed with personality and even knowledge, which can be associated from the psychological perspective with anthropomorphism. This is found in other statements like "How does ChatGPT know how to formulate learning goals?" (S33) or "if the AI does know everything" (S19). A detailed description of the Large Language Model was not found.

## V. DISCUSSION

The study aimed to engage students in critically reflecting on AI-generated content within a didactic context and to help them develop media literacy. Although prompt engineering can optimize AI output, it falls outside the scope of this study; instead, the focus is on evaluating students' critical reflection on the tasks. It is worth noting that students submitted their questions to ChatGPT exactly as they were phrased in the homework, without any modifications or additional prompts. During discussions, they confirmed that they used the same wording from the tasks provided.

The **first research question** investigated how physics teacher students evaluate the quality of ChatGPT3.5-generated content for didactic tasks, such as developing context-oriented teaching in acoustics and formulating learning goals. The students effectively applied their pedagogical knowledge in evaluating learning goals, particularly recognizing ChatGPT's deficiencies in using operational verbs. However, they struggled to apply their knowledge of context-oriented teaching, despite this being covered in their introductory physics didactics course. These findings suggest the need for greater emphasis on context-oriented teaching in the course curriculum.

The **second research question** investigated how physics teacher students evaluated ChatGPT 3.5-generated content for didactic tasks using given criteria. The task involved developing teaching-learning sequences on energy and power for 7th-grade students. Students applied the given criteria – clarity, correctness, appropriateness, and precision – to evaluate ChatGPT's content. They identified several deficiencies: a lack of clarity due to vague terminology, incorrect content, lack of appropriateness by failing to consider the students' grade and prior knowledge, and lack of precision, as the teaching-learning sequences were superficial and lacked a logical structure.

Additionally, when asked about the usefulness of the criteria for evaluating ChatGPT's answers, most students (18 out of 20) found the criteria helpful and appropriate. The following statement illustrates this perspective:

S24 (in written form): *"The criteria list helped a lot because otherwise I wouldn't have known how to evaluate ChatGPT's answer."*

The **third research question** investigated how engaging physics teacher students in evaluating ChatGPT3.5-generated content impacts their perception of ChatGPT's helpfulness and its answer quality. The results showed a slight change in the students' perception of the answer quality and helpfulness of ChatGPT between the pre- and post-questionnaire ratings. The histogram (Fig. 2) shows a slight decrease in ratings, with the average rating of response quality dropping from 3.16 to 2.66. The evaluation of ChatGPT's helpfulness also decreased, indicating that students were more critical of the AI's capabilities. Although these changes were not statistically significant, they indicate an initial optimism that decreased through practical application.

The qualitative analysis of students' answers offered deeper insight into their evaluations. Students who reported an **increase in perceived helpfulness** (see Fig. 3) appreciated ChatGPT's ability to provide quick answers, support brainstorming, and suggest step-by-step teaching-learning sequences for topics like energy and power. They specifically highlighted ChatGPT's ability to provide quick answers as a key factor in justifying its helpfulness.

Conversely, students who reported a **decrease in perceived helpfulness** emphasized the importance of clarity, correctness, appropriateness and precision in educational content. Those who rated the **quality of ChatGPT's answers lower** also emphasized these criteria and used them to critically evaluate and discuss the quality of responses.

Interestingly, a small group of students who maintained or improved their ratings of ChatGPT's answer quality revealed signs of **confirmation bias**. Although they acknowledged deficiencies in the answers, they were impressed by the ChatGPT's ability to provide examples and solve common tasks rapidly. This tendency to overlook shortcomings in favor of positive aspects could be due to previous experience with ChatGPT. Such biases may hinder critical evaluation, suggesting that students' previous positive impressions may influence their critical reflection of the content generated by the AI.

Furthermore, three perspectives on the further use of ChatGPT were observed in student' discussions: extreme negative, extreme positive, and balanced.

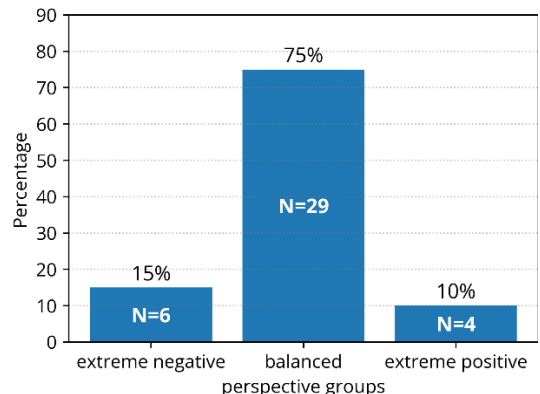

FIG 4. Distribution of students across the three perspective-groups on the future use of ChatGPT.

Fig. 4 shows the number and the percentage of students in each perspective-group. The **extreme negative perspective** rejected the further use of ChatGPT, viewing it as a time-consuming effort. This perspective was mostly observed among students with no prior experience using ChatGPT. The following student's statement illustrates this perspective well: "*I wouldn't use it. So if it really stays with the quality as I've experienced it now, nope. Because I'd have to sit down there again to do more work ...*" (S25).

The **extreme positive perspective** favored the further use of ChatGPT, despite recognizing deficiencies in its answers for didactic tasks. This view was observed among students who had prior experience with ChatGPT3.5 and were more impressed by its capabilities in general tasks.

Engaging this group in a criterion-based evaluation and their almost critical reflection on the answers was not reflected in their perception of the quality of ChatGPT's answers and helpfulness in the post-questionnaire. The following student's statement illustrates this perspective well: "*On the subject of clarity, I think the wording is clear, it's just superficial, but you understand what the AI means.*" (S34)

As previously discussed, this may be explained by **confirmation bias** [17], where the perception of one positive trait (ChatGPT's general capabilities) influences the evaluation of other, unrelated traits (such as the quality of its specific answer).

The **balanced perspective**, however, emphasized the importance of critically reflecting on ChatGPT's output. Students with this view appreciated its role as an assistive tool but also recognized the need for critical evaluation of ChatGPT-generated content. They highlighted the value of discussing its pros and cons, especially with high school students, to ensure a more thoughtful and informed use of the technology.

S07: "*I found it important to discuss advantages and disadvantages of using ChatGPT with students. Because students will use it. And I definitely see it as my responsibility as a teacher that I have to address this ....*"

The **fourth research question** investigated how physics teacher students perceive and interpret the functionality of ChatGPT. The results showed a lack of correct understanding of the functionality of ChatGPT. The functionality was interpreted predominantly with comparisons to a database or an optimized search engine. Thorough, detailed, and accurate descriptions of ChatGPT's functionality were not found. However, this was expected, as there was no explicit task to make such descriptions. Students used also typical anthropomorphisms ("he knows") [18,19], which shows that the way of how to talk and write about ChatGPT and other bots is dominated by the Chat-type interaction, which comes along with certain risks [19,20].

## VI. CONCLUSIONS

The study introduced a criterion-based evaluation approach, where physics teacher students evaluated ChatGPT 3.5-generated content on didactic tasks using specific criteria. Students' critical reflection on the ChatGPT-generated content using criteria caused a shift in their perspective on ChatGPT's answer quality and also made students more aware of its limitations. Despites this awareness, some students showed confirmation bias, where their positive prior experiences with ChatGPT influenced their evaluations. This highlights the importance of developing students' media literacy, enabling them to adopt a balanced approach, rather than completely rejecting digital tools or allowing confirmation bias to influence their evaluation. Developing this skill addresses the application-related aspect of media literacy.

Furthermore, findings from RQ4 highlights students' challenges in understanding ChatGPT's functionality, emphasizing the technological aspect of media literacy. The majority of students had only a vague or incorrect understanding of its functionality as an artificial neural network, emphasizing the need to address this aspect in teacher training programs. Overall, using a criterion-based evaluation approach on ChatGPT-generated content promotes students' critical reflection and promises development of media literacy in students. This study therefore suggests integrating this approach into teacher training programs, as AI tools like ChatGPT are increasingly becoming part of students' lives. Doing so can strengthen their critical thinking and encourage mindful use of technology.

Moreover, ChatGPT-generated content can be used to evaluate students' pedagogical content knowledge, as shown in the findings from RQ1. This can inform strategies in teacher training that not only cover theoretical concepts but also emphasize real-world application.

Finally, insights into how students evaluated ChatGPT's helpfulness and the presence of confirmation bias in evaluating its answer quality can guide the development of targeted scenarios or vignettes for university and high school students, promoting critical thinking and media literacy.

## ACKNOWLEDGMENTS

Open access funding provided by the Open Access Publishing Fund of the TU Dresden, funded by the DFG and supported by the Sächsische Landes- und Universitätsbibliothek Dresden (SLUB).